\documentclass[twocolumn]{article}

\usepackage{algorithm}
\usepackage{algpseudocode}
\usepackage{booktabs}
\usepackage{endnotes}
\usepackage{enumitem}
\usepackage{epsfig}
\usepackage{float}
\usepackage{graphicx}
\usepackage{hyperref}
\usepackage{multirow}
\usepackage{subfigure}

\begin{document}

\title{FGo: A Directed Grey-box Fuzzer with Probabilistic Exponential \textit{cut-the-loss} Strategies}

\author{
    {\rm Harvey Lau}
}

\maketitle

\subsection*{Abstract}

Traditional coverage grey-box fuzzers perform a \textit{breadth-first search} of the \textit{state space} of Program Under Test (PUT). This aimlessness wastes a lot of computing resources. Directed grey-box fuzzing focuses on the target of PUT and becomes one of the most popular topics of software testing. The early termination of unreachable test cases is a method to improve directed grey-box fuzzing. However, existing solutions have two problems: firstly, reachability analysis needs to introduce extra technologies (e.g., static analysis); secondly, the performance of reachability analysis and auxiliary technologies lack versatility.

We propose FGo, a probabilistic exponential \textit{cut-the-loss} directed grey-box fuzzer. FGo terminates unreachable test cases early with exponentially increasing probability. Compared to other technologies, FGo makes full use of the unreachable information contained in iCFG and doesn‘t generate any additional overhead caused by reachability analysis. Moreover, it is easy to generalize to all PUT. This strategy based on probability is perfectly adapted to the randomness of fuzzing.

The experiment results show that FGo is 106\% faster than AFLGo in reproducing crashes. We compare multiple parameters of probabilistic exponential \textit{cut-the-loss} algorithm and analyze them in detail. In addition, for enhancing the interpretability of FGo, this paper discusses the difference between the theoretical performance and the practical performance of probabilistic exponential \textit{cut-the-loss} algorithm.

\section{Introduction}


Grey-box fuzzing is a popular automated vulnerability detecting technique. It takes improving code coverage as the goal and performs a \textit{breadth-first search} on PUT. Because the more code coverage, the higher the possibility of discovering bugs. However, the proportion of code containing bugs is very small. This means that lots of computing resource used to improve code coverage are wasted. To solve this problem, directed grey-box fuzzing appeared. Directed grey-box fuzzer focuses on the target of PUT and tilts computing resource towards it, which performs \textit{depth-first search} on PUT. From the perspective of \textit{state space}, the \textit{states} of coverage grey-box fuzzing is all paths while the \textit{states} of directed grey-box fuzzing is some specific paths.

The essence of fuzzing is to obtain the information of PUT through test cases. As far as the application scenario of fuzzing is concerned, the information of PUT we need is bugs. Grey-box fuzzing often uses the known information of PUT to increase the speed of searching \textit{states} or reduce \textit{state space}. The existing directed grey-box fuzzers obtain the call graph (CG) and control flow graph (CFG) of PUT through static analysis and set the target of PUT through known vulnerabilities\cite{bohme2017directed}. Firstly, directed grey-box fuzzer extracts CG and CFG from the source code of PUT; secondly, combined with the target of PUT, directed grey-box fuzzer pre-computes all function distance and basic block (BB) distance with the target functions and the target BBs; thirdly, directed grey-box fuzzer uses the feedback information obtained at runtime to reduce the distance of test cases and approach the target until the corresponding vulnerability is triggered.

Improvements to directed grey-box fuzzers can be divided into two categories: improving the framework of directed grey-box fuzzing before getting test cases\cite{chen2018hawkeye, du2022windranger} and improving the execution process of PUT after getting test cases\cite{zong2020fuzzguard, huang2022beacon}. As shown in Figure 1.a, if we improve the framework of directed grey-box fuzzing before getting test cases, such as distance, power schedule, and mutation strategy, it is similar to putting a stone in the upper reaches of a river. Although this improvement can tilt the computing resource towards the target, its effect is very limited. As shown in Figure 1.b, if we improve the execution process of PUT after getting test cases, such as early termination, it is similar to building a dam in the middle reaches of a river. This improvement can greatly increase the efficiency of directed grey-box fuzzing.

The key information to improve the execution process of PUT after getting test cases is reachability. FuzzGuard\cite{zong2020fuzzguard} divides test cases into reachable and unreachable by neural network. Through backward interval analysis, BEACON\cite{huang2022beacon} pre-inserts the conditions which correspond to reaching the target. They have two commonalities: firstly, the reachability analysis needs to introduce other technologies; secondly, the effect of the reachability analysis depends on specific PUTs. FuzzGuard\cite{zong2020fuzzguard} predicts reachability by pre-trained neural network. Moreover, vulnerabilities are various and therefore neural network trained on a few specific PUT lack versatility. Similarly, BEACON requires static analysis and the preconditions are only numerical conditions\cite{huang2022beacon}.

We found that existing directed grey-box fuzzers do not make full use of control flow information (iCFG). In fact, the distances pre-computed by AFLGo\cite{bohme2017directed} implicitly imply the reachability information. From the perspective of reachability, the path of a test case is divided into two segments: in iCFG, the functions and BBs in the previous segment are reachable to the target while the functions and BBs in the latter segment are unreachable to the target. Therefore, when the execution process of a test case enters the unreachable segment, it should be terminated early. However, existing directed grey-box fuzzers do not take this into consideration and cannot \textit{cut-the-loss} in time.

This paper focuses on the known iCFG to terminate unreachable test cases early. We propose FGo, a probabilistic exponential \textit{cut-the-loss} directed grey-box fuzzer. FGo terminates unreachable test cases early with exponentially increasing probability. Without early termination, it prevents false positives of unreachable test cases. In the case of early termination, it saves lots of computing resource. Compared to directed grey-box fuzzers introducing other technologies, FGo makes full use of the reachability information contained in the known iCFG and doesn't generate any additional overhead caused by reachability analysis. Moreover, it is easy to generalize to all PUTs. This strategy based on probability is perfectly adapted to the randomness of fuzzing. We implement the prototype of FGo on AFLGo.

The contributions of this paper are as follows:

\begin{itemize}
    \item Without introducing any additional overhead caused by reachability analysis, we propose probabilistic exponential \textit{cut-the-loss} strategies to terminate unreachable test cases early. This idea can be generalized to all directed grey-box fuzzers and PUTs.
    \item We extend AFLGo to implement FGo and publish its source code.
    \item We evaluate FGo on real programs and analyze experiment results in detail.
    \item We study the theoretical performance of probabilistic exponential \textit{cut-the-loss} algorithm and analyze its difference from the practical performance.
\end{itemize}

To facilitate the research in this field, we open source FGo at \url{https://github.com/harvey-lau/fgo}.

\section{Background}

\subsection{Directed Grey-box Fuzzing}

According to the degree of dependence on the known information of PUT, fuzzing can be divided into black-box, white-box and grey-box\cite{manes2019art}. Black-box fuzzing\cite{beizer1995black} considers PUT as a black box that no information about PUT is known before fuzzing and only the output corresponding to the input can be observed. White-box fuzzers\cite{ostrand2002white} generate test cases based on analyzing PUT and their runtime information. Grey-box fuzzing\cite{zalewski2014american} performs lightweight static analysis on PUT or collects some runtime information of test cases to guide the mutation of test cases. Black-box fuzzing is not effective enough while the effectiveness of white-box fuzzers pays the cost of efficiency\cite{bohme2015probabilistic}. Grey-box fuzzing strikes a balance between effectiveness and efficiency and therefore becomes the most widely used fuzzing method.

At first, grey-box fuzzers aim to improve code coverage\cite{bohme2016coverage}. Because the higher the code coverage, the greater the possibility of discovering vulnerabilities. However, the path corresponding to the vulnerability accounts for a very small proportion of code\cite{wang2020progress}. It means a large number of computing resource used to improve code coverage can not improve vulnerability coverage. Therefore, blindly improving code coverage is very inefficient. The empirical analysis of coverage-based fuzzer benchmarking shows that the relationship between of the coverage achieved and the number of vulnerabilities found is strong correlation but isn't strong agreement\cite{bohme2022reliability}.

Directed fuzzing focuses on the target of PUT and tilts computing resource towards the target, rather than trying to cover all paths. In the beginning, directed fuzzing use symbolic execution to generate inputs that triggered specific paths. It is often based on a symbolic execution engine, such as KLEE\cite{cadar2008klee}. However, directed symbolic execution relies on program analysis and constraint solving which bear unacceptable time overhead. Compared to directed symbolic execution, directed grey-box fuzzing is not only suitable for large programs, but also has strong flexibility.

\subsection{The Distance and Reachability of Test Cases}

AFLGo is the first directed grey-box fuzzer\cite{bohme2017directed}. It regards the process of reaching the target of PUT as an \textit{optimization problem} and considers the distance between a test case and the target as \textit{loss function}. AFLGo defines \textit{distance} in three steps.

\textit{Function-level distance} defines the distance between a function and all target functions. Firstly, the distance of two functions is defined as the shortest path length in CG; Secondly, the distance between a function and a set of functions is defined as the harmonic mean of the distances between a function and each function.

\textit{BB-level distance} defines the distance between a BB and all target BBs. Different situations are discussed as follows:

\begin{enumerate}[label=\alph*)]
    \item If the BB belongs to the BB set of the target, the distance is zero.
    \item If the function of the BB and the function set of the target are reachable, the BB and the BB set of the target is reachable. The \textit{BB-level distance} is defined as a multiple of the \textit{function-level distance}.
    \item If the BB has no direct reachability relationship with the BB set of the target, the \textit{BB-level distance} is defined as the harmonic mean of the distance between the BB and each BB in the same CFG.
\end{enumerate}

\textit{Normalized distance} defines the distance between a test case and the target. We can regard the path of a test case as a series of BBs. Based on the distance between a BB and all target BBs defined above, the distance between a test case and the target is defined as the arithmetic mean of the distance between each BB of a test case and the target BBs. In order to make the \textit{maximum distance} not change with the scale of PUT, based on the \textit{minimum distance} and the \textit{maximum distance}, the distance of a test case is normalized to $[0, 1]$.

So, how does AFLGo handle unreachable test cases?

\begin{enumerate}[label=\alph*)]
    \item In \textit{function-level distance}, The distances of unreachable functions are undefined.
    \item In \textit{BB-level distance}, if the BB is unreachable to the BB set of the target, since its distance is based on \textit{function-level distance}, The distances of unreachable BBs are also undefined.
    \item So is \textit{test case-level distance}.
\end{enumerate}

Let us turn to the implementation of AFLGo\cite{githubGitHubAflgoaflgo}. At compile time, AFLGo initializes the distance between a BB and the target as $-1$. If the initial distance isn't updated by a positive, the BB doesn't participate in the instrumentation process of calculating the final distance. At runtime, AFLGo initializes the distance between a test case and the target as $-1$. If the initial distance isn't updated by a positive, the test case doesn't participate in the power schedule based on simulated annealing.

Based on reachability, FuzzGuard and BEACON \textit{cut-the-loss}. However, both of their reachability analysis are enabled by auxiliary technologies. FuzzGuard needs to pre-train a classification neural network through test cases and their reachability information\cite{zong2020fuzzguard}. BEACON needs to extract numerical conditions from the source code of PUT through backward interval analysis\cite{huang2022beacon}.

\subsection{Exponential Backoff Algorithm}

Exponential backoff algorithms\cite{wiki:Exponential_backoff} are used in control systems to reduce the rate of responding to adverse events. For example, if a client cannot connect to the server, it will try again after 1ms; if it fails again, it will try again after 2ms, 4ms, 8ms, ..., and so on.

The mathematical model of reducing the rate of responding to adverse events is an exponential function:

$$
t = c^k, k = 0, 1, 2, ...
$$

\begin{itemize}
    \item $t$ represents the delay between adverse events.
    \item $c$ represents the multiple of each delay.
    \item $k$ represents the number of adverse events.
\end{itemize}

When $c = 2$, exponential backoff algorithm is also known as binary exponential backoff algorithm.

In this paper, the backoff of adverse events is replaced by the \textit{cut-the-loss} of test cases.

\section{Motivating Example}

\begin{figure*}[h]
    \centering
    \setcounter {subfigure} {0} a){
        \includegraphics[width=0.305\textwidth]{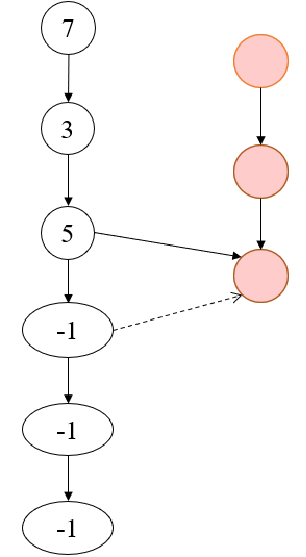}}
    \setcounter {subfigure} {0} b){
        \includegraphics[width=0.295\textwidth]{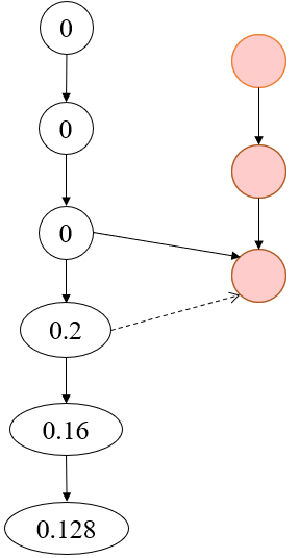}}
    \caption{Motivating Example. a) The left side is a test case and the right side is the target of PUT. The number of the left node represents the distance between the node and the last node of the target where $-1$ means unreachable. b) The left side is a test case and the right side is the target of PUT. The number of the left node represents the \textit{cut-the-loss} probability of the node.}
    \label{fig2} 
\end{figure*}

As shown in Figure 2, let white nodes represent the BBs of a test case and red nodes represent the BBs of the target. In Figure 2.a, the number of white nodes represents the distance between this BB and the last BB of the target. In Figure 2.b, the number of white nodes represents the \textit{cut-the-loss} probability of this BB.

Firstly, assuming that it is reachable between a test case and the first sub-target, then it is reachable between a test case and the third sub-target.

Secondly, assuming that it is reachable between the second BB of a test case and the last sub-target, then it is reachability between the first BB of a test case and the last sub-target. It is worth mentioning that the distance of the first BB doesn't necessarily increase on the distance of the second BB, because the path of going through the second BB may be one of several paths between the first BB and the target.

Thirdly, assuming that it is unreachable between the fourth BB of a test case and the last sub-target, then it is unreachable between the fifth BB of a test case and the last sub-target.

Based on the above analysis, it can be known that:

\begin{enumerate}[label=\roman*]
    \item The reachability between a test case and the target is equivalent to the reachability between a test case and the last sub-target.
    \item A test case transfers from reachable to unreachable (if such transfer exists).
\end{enumerate}

Because the first conclusion about reachability involves the completeness of iCFG, it is not the point of this paper. For the second conclusion, we use the idea of exponential backoff algorithm to calculate the \textit{cut-the-loss} probability of each unreachable BB.

We know that a test case can be divided into reachable segment and unreachable segment. As shown in Figure 2.b, if we set exponential \textit{cut-the-loss} probability to $0.2$, the probability that the test case is terminated at the first unreachable BB is $0.2$, the probability that the test case is terminated at the second unreachable BB is $(1 - 0.2) \cdot 0.2 = 0.16$, the probability that the test case is terminated at the third unreachable BB is $(1 - 0.2)^2 \cdot 0.2 = 0.128$ , the probability that the test case is terminated at the $i$-th unreachable BB is $(1 - 0.2)^{i - 1} \cdot 0.2$, ..., and so on.

\section{Methodology}

\begin{figure}[h]
    \begin{center}
    \includegraphics[width=0.45\textwidth]{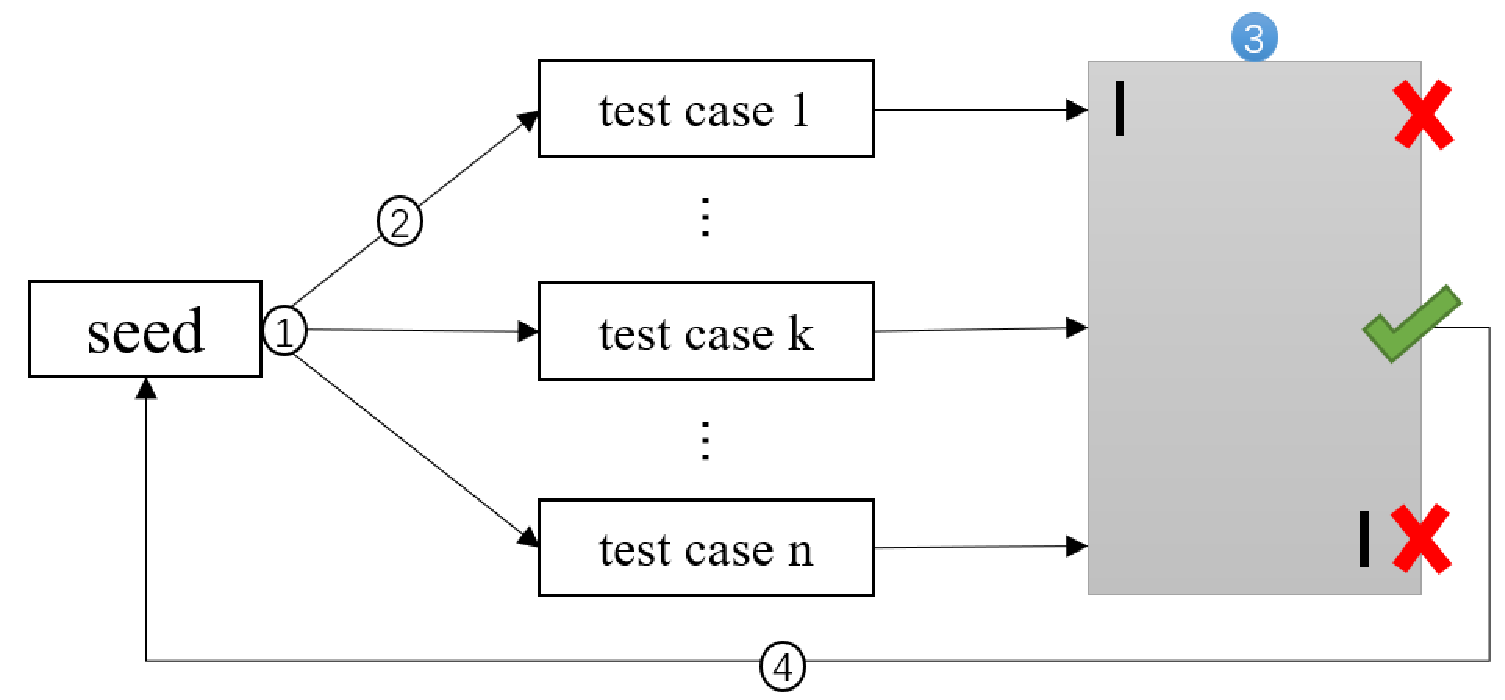}
    \end{center}
    \caption{The Fuzzing Loop of A Seed.}
    \label{fig3}
\end{figure}

The workflow of FGo is shown in Figure 3. We abstract the process of fuzzing a seed into four steps.

\begin{enumerate}
    \item Firstly, according to its importance, a seed is assigned some given computing resource (e.g., the number of mutations $n$).
    \item Secondly, one seed is mutated to $n$ test cases in the light of mutation strategy.
    \item Thirdly, the reachable test cases are executed normally while the unreachable test cases are terminated early. For example, if the BBs of test case $1$ almost are unreachable, test case $1$ is terminated at the beginning of execution process; if the BBs of test case $n$ almost are reachable, test case $n$ is terminated at the end of execution process; if test case $k$ is reachable, it is executed normally.
    \item Fourly, If a test case is interesting, it will be added into seed pool.
\end{enumerate}

\subsection{Probabilistic Exponential \textit{cut-the-loss} Algorithm}

\begin{algorithm}
    \caption{Probabilistic Exponential \textit{cut-the-loss}}
    \label{alg1}
    \hspace*{\algorithmicindent} \textbf{Input:} test case \textit{t} \\
    \hspace*{\algorithmicindent} \textbf{Output:} void
    \begin{algorithmic}[1]
        \For {BB in test case $t$}
            \If{$\textit{BB-distance} < 0$}
                \State srand(time(0));
                \If {$rand()\ mod\ 10 + 1 > (1 - p) * 10$}
                    \State break;
                \EndIf
            \EndIf
        \EndFor
    \end{algorithmic}
\end{algorithm}

Algorithm 1 represents the execution of a test case with a probabilistic exponential \textit{cut-the-loss} algorithm. \texttt{srand(time(0))} initializes a random number seed; \texttt{rand()} generates a random integer; $p$ represents exponential \textit{cut-the-loss} probability. For each BB of a test case, if its distance is $-1$, \texttt{rand() mod 10 + 1} gets a random integer between $1$ and $10$. If the integer is greater than $(1 - p) \cdot 10$, the test case will be terminated early. As a result, probabilistic exponential \textit{cut-the-loss} algorithm achieves the early termination of test case $t$ at each unreachable BB with exponential \textit{cut-the-loss} probability $p$.

\subsection{\textit{True Positive}}

We refer to unreachable test cases with early termination as \textit{true positive} test cases. BEACON shows that more than 95\% of unreachable test cases are fully executed\cite{huang2022beacon}. If they are terminated early, the performance of directed grey-box fuzzers will be greatly improved.

\begin{figure}[h]
    \centering
    \setcounter {subfigure} {0} a){
        \includegraphics[width=0.45\textwidth]{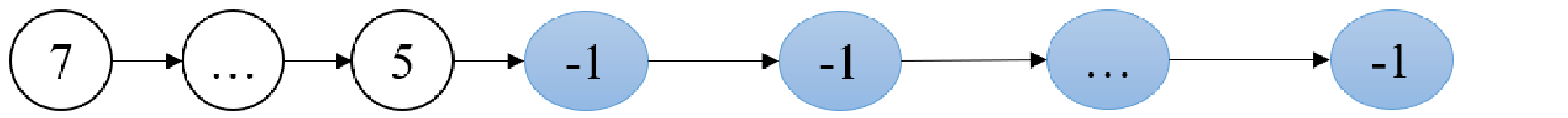}}
    \\
    \setcounter {subfigure} {0} b){
        \includegraphics[width=0.45\textwidth]{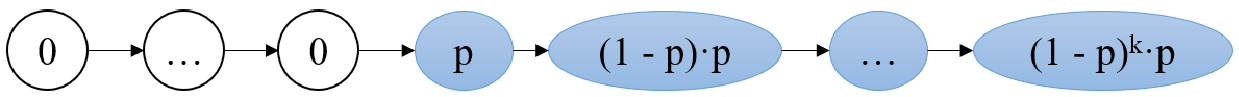}}
    \caption{The distance and \textit{cut-the-loss} probability of a test case's each BB. White nodes represent reachable BBs while blue nodes represent unreachable BBs. a) The number of each node represents the distance of each BB. b) The number of each node represents the \textit{cut-the-loss} probability of each BB.}
    \label{fig1}
\end{figure}

As shown in Figure 4, it represents the distance and \textit{cut-the-loss} probability of a test case's each BB. White nodes represent reachable BBs while blue nodes represent unreachable BBs. The computing resource saved by the test case is terminated early depends on the number of unreachable BBs. As shown in Figure 4.b, if a test case is terminated early at the first unreachable BB, the computing resource for executing the subsequent $k$ unreachable BBs is saved. From the perspective of \textit{static analysis}, the \textit{cut-the-loss} probability at each unreachable BB is not the same. For example, the \textit{cut-the-loss} probability at the second unreachable equals the product of the probability of not early termination at the first unreachable BB and the probability of early termination at the second unreachable BB. In summary:

\begin{enumerate}[label=\roman*]
    \item The more unreachable BBs, the greater the probability that the test case is terminated early and the more computing resource is saved.
    \item The later unreachable BBs, the lower the probability that the BB is executed.
\end{enumerate}

We conduct an theoretical analysis in Section 7.

\subsection{\textit{False Positive}}

We refer to reachable test cases with early termination as \textit{false positive} test cases. What we don't want to happen is that \textit{false positive} test cases are terminated early.

What if FGo terminate unreachable test cases early with exponential \textit{cut-the-loss} probability $p = 1$? Assuming that iCFG can completely represent PUT and the distance information completely corresponds to the reachability of every BB (e.g., non-negatives represent reachable while negatives represent unreachable), let $p = 1$ can maximize the performance of FGo. However, that function-level distance or BB-level distance equals $-1$ doesn't mean certainly unreachable\cite{chen2018hawkeye}. Therefore, there are false positive test cases with early termination when exponential \textit{cut-the-loss} probability $p$ equals $1$. We need $1 - p$ to prevent reachable test cases from being terminated. Furthermore, in the early stages of fuzzing, if FGo terminate test cases early too frequently, this will significantly affect the original feedback mechanism of AFL and make directness fall into a local optimum. With a high-level view, the exponential \textit{cut-the-loss} probability $p$ is a trade-off between \textit{true positives} and \textit{false positives}.

\section{Implementation}

We implemented a prototype of FGo in \texttt{C} and \texttt{C++} as an extension of AFLGo. At runtime, FGo calls a function \texttt{noway()} to realize probabilistic exponential \textit{cut-the-loss} algorithm. In \texttt{noway()}, we use a random integer between $1$ and $10$ to decide whether a test case is terminated early with a give exponential \textit{cut-the-loss} probability $p$ or not. Similar to BEACON\cite{huang2022beacon}, termination is achieved by \texttt{assert(false)}.

\section{Evaluation}

\textbf{Benchmarks.} We used LibMing version 0.4.7\cite{githubGitHubLibMingLibMing} containing 4 different categories of crashes which correspond to different function call stacks. BEACON\cite{huang2022beacon} shows that LibMing version 0.4.7 contain 4 vulnerabilities assigned CVE ID and they can be reproduced in 6 hours. Because the CVE descriptions of them are too vague to set the target of LibMing version 0.4.7, we first test LibMing version 0.4.7 through AFL\cite{githubGitHubGoogleAFL}. According to our analysis, the crashes of LibMing version 0.4.7 found by AFL are divided into 4 categories, as shown in Table 1. Then, we consider the most common function call stacks of the 4 categories of crashes as 4 different targets of LibMing version 0.4.7. We determine which bug a test case triggers by the feature of crash.

\begin{table}[h]
    \caption{The Crash ID of crashes and their features of LibMing version 0.4.7.}
    \label{tab1}
    \renewcommand\arraystretch{1.5}
    \resizebox{\columnwidth}{!}{
    \begin{tabular}{c|c|c}
    \hline
    Program                         & Crash ID & Feature        \\ \hline
    \multirow{4}{*}{LibMing version 0.4.7} & min1     & parseABC       \\ \cline{2-3} 
                                    & min2     & outputSWF      \\ \cline{2-3} 
                                    & min3     & readBytes      \\ \cline{2-3} 
                                    & min4     & parseSWF\_RGBA \\ \hline
    \end{tabular}}
\end{table}

\textbf{Fuzzers.} We compare FGo with AFLGo.

\begin{itemize}
    \item AFLGo, the first and most widely used directed grey-box fuzzer.
    \item FGo, the tool proposed in this paper.
\end{itemize}

\textbf{RQs.} We conducted all experiments to answer following questions:

\begin{itemize}
    \item Time to Exposure (TTE): does FGo outperform AFLGo in reproducing crashes?
    \item The Different parameters of probabilistic exponential \textit{cut-the-loss} algorithm: how do the parameters of probabilistic exponential \textit{cut-the-loss} algorithm affect its performance?
\end{itemize}

We used the seeds in the GitHub repository of AFLGo and repeated every experiment 7 times with a timeout 6 hours. We set the time-to-exploitation $t_x$ to 2 hours. The command of fuzzing is:

\texttt{/path/to/fgo -m none -z exp -c 120m -i in -o out -t 5000+ ./util/swftophp @@}

We conducted all experiments on a computer with Intel(R) Core(TM) i7-10700 CPU @ 2.90GHz with 16 cores, 16GB memory, and Ubuntu 20.04.1 LTS.

\subsection{Time to Exposure (TTE)}

\begin{table}[h]
    \caption{The TTE Results of AFLGo and FGo (s).}
    \label{tab2}
    \renewcommand\arraystretch{1.5}
    \resizebox{\columnwidth}{!}{
    \begin{tabular}{ccccc}
    \hline
    Program                                       & Crash ID                         & AFLGo   & FGo         & Speedup \\ \hline
    \multicolumn{1}{c|}{\multirow{4}{*}{LibMing}} & min1                      & 7706.14 & 5487.14     & 1.40    \\
    \multicolumn{1}{c|}{}                         & min2                      & 1131.57 & 246.00      & 4.60    \\
    \multicolumn{1}{c|}{}                         & min3                      & 647.57  & 3322.14 (6) & 0.19    \\
    \multicolumn{1}{c|}{}                         & min4 & T.O.    & T.O.        & T.O.    \\ \hline
    \end{tabular}}
\end{table}

Compared to AFLGo, the performance of FGo at exponential \textit{cut-the-loss} probability $p = 0.1$ is improved by an average of 106\% as shown in Table 1.

\begin{itemize}
    \item For \texttt{min1}, it is a crash difficult to be reproduced. The crash number of AFLGo triggering this crash ranges from 18 to 31. And the TTE fluctuates between 5573s and 13258s in 7 repeated experiments. In contrast, FGo can filter out lots of irrelevant crashes (e.g., the crashes of \texttt{min2}), which the minimum crash number of triggering \texttt{min1} is only 2. In this situation, it only took 440s for FGo to reproduce it. Because the TTE of \texttt{min1} is relatively long, its reproduction is more dependent on multiple specific mutations.Therefore, the increased performance of FGo is not high, about 40\%.
    \item For \texttt{min2}, this is the crash to be reproduce most easily. The crash number of AFLGo triggering the crash is stably 1. Its TTE is relatively small. Therefore, the interference of early termination in the coverage feedback mechanism of AFL is almost negligible. In such cases, the effect of FGo's probabilistic exponential \textit{cut-the-loss} algorithm is very significant, increasing the performance by an average of 360\%.
    \item For \texttt{min3}, it is a crash with moderate difficulty in reproduction. The crash number of AFLGo triggering this crash is about 7. In terms of TTE, FGo performs poorly. The maximum TTE of FGo was as big as 17570s. The possibly reason why AFLGo outperformed FGo is that probabilistic exponential \textit{cut-the-loss} algorithm has caused FGo to fall into a local optimum.
    \item For \texttt{min4}, neither AFLGo nor FGo can reproduce this crash.
\end{itemize}

\subsection{The Different parameters of probabilistic exponential \textit{cut-the-loss} algorithm}

To study the performance of different \textit{cut-the-loss} probability, we respectively tested LibMing version 0.4.7 with $p = 0.1$, $p = 0.2$, and $p = 0.4$.

\begin{table}[h]
    
    \caption{The Different Parameters results of FGo (s).}
    \label{tab3}
    \renewcommand\arraystretch{1.5}
    \resizebox{\columnwidth}{!}{
    \begin{tabular}{ccccc}
    \hline
    Program                                       & Crash ID                         & FGo 0.1     & FGo 0.2     & FGo 0.4     \\ \hline
    \multicolumn{1}{c|}{\multirow{4}{*}{LibMing}} & min1                      & 5487.14     & 5419.33 (6) & 7076.40 (5) \\
    \multicolumn{1}{c|}{}                         & min2                      & 246.00      & 383.14      & 1325.33 (6) \\
    \multicolumn{1}{c|}{}                         & min3                      & 3322.14 (6) & 4039.29     & 5393.33 (6) \\
    \multicolumn{1}{c|}{}                         & min4 & T.O.        & T.O.        & T.O.        \\ \hline
    \end{tabular}}
\end{table}

As shown in Table 3, $p = 0.1$ outperforms $p = 0.2$ and $p = 0.4$. Because the larger the value of $p$, the lower the stability of probabilistic exponential \textit{cut-the-loss} algorithm. Although $p = 0.1$ seems very small, it only represents the probability of early termination at the first unreachable BB. The probability of a test case being terminated is $1 - (1 - p)^u$ where $u$ represents the number of unreachable BBs. For example, when $p = 0.1$, $u = 10$, the probability of a test case being terminated is 0.65; when $p = 0.1$, $u = 20$, the probability of a test case being terminated is 0.88.

\begin{table*}[h]
\caption{The Explanation of Symbols}
\label{tab4}
\renewcommand\arraystretch{1.5}
\begin{tabular}{c|c}
\hline
Symbols        & Explanation                                                                                           \\ \hline
$p$           & exponential \textit{cut-the-loss} probability                                                                     \\
$t_1$         & the time overhead of fuzzing loop                                                                     \\
$t_2$         & the time overhead of execution process                                                                \\
$t_2'$        & the time overhead of execution process with probabilistic exponential \textit{cut-the-loss} algorithm \\
$T$           & $T = t_1 + t_2$                                                                                       \\
$N$           & the number of test cases                                                                              \\
$r_i$         & the number of test case $i$'s reachable BBs                                                           \\
$u_i$         & the number of test case $i$'s unreachable BBs                                                         \\
$\overline r$ & $\frac 1 N \sum\limits_{i = 1}^{N} r_i$                                                                    \\
$\overline u$ & $\frac 1 N \sum\limits_{i = 1}^{N} u_i$                                                                     \\
$s$           & $s = 1 - \frac {t_2'} {t_2}$                                                                              \\
$N$           & the number of test cases                                                                              \\ \hline
\end{tabular}
\end{table*}

\section{Discussion}

\subsection{The Theoretical Analysis}

Based on the definitions in Table 4, the ratio of the time overhead of execution process $t_2$ and the time overhead of execution process with probabilistic exponential \textit{cut-the-loss} algorithm $t_2'$ equals: 

\begin{equation}
s = \sum\limits_{i = 1}^{\overline u} (1 - p)^{i - 1} \cdot p \cdot \frac {\overline u + 1 - i} {\overline r + \overline u}
 \label{eq1}
\end{equation}

$s$ represents the proportion of time overhead reduced by probabilistic exponential \textit{cut-the-loss} algorithm, where $s \in [0, 1]$. Therefore, the theoretical value of speedup is:

\begin{equation}
I = \frac {T} {T'} = \frac {t_1 + t_2} {t_1 + t_2'} = \frac {t_1/t_2 + 1} {t_1/t_2 + 1 - s}
 \label{eq2}
\end{equation}

Obviously, $\frac {dI} {ds} > 0$ and therefore the sign of $\frac {dI} {dp} = \frac {dI} {ds} \cdot \frac {ds} {dp}$ depends on the sign of $\frac {ds} {dp}$.

For better performance, FGo should terminate unreachable test cases as early as it can. Therefore, $\frac {ds} {dp} > 0$ and $\frac {dI} {dp} > 0$. It means that the larger $p$ is, the higher FGo performs.

\subsection{Future Work}

As analyzed above, we should choose $p$ as large as possible. Then, why does $p = 0.1$ outperform $p = 0.2$ and $p = 0.4$? There are two major reasons: incomplete pre-computed distances and the interference of early termination in the coverage feedback mechanism of AFL. We can derive two future works from these reasons.

Firstly, we should reconstruct the definition and calculation of distance to complete pre-computed distance files. Secondly, we can transfer from exponential model into a sophisticated model to reduce or eliminate its negative impact. For example, FGo does nothing in exploration phrase and runs an exponential \textit{cut-the-loss} algorithm in exploitation phrase. This help FGo to avoid falling into a local optimum.

\section{Related Work}

Let us summarize the related works mainly in chronological order.

\textbf{Coverage-based Grey-box Fuzzing.} AFL is one of the most classic coverage-based grey-box fuzzer. Lots of works base on AFL and optimize the framework of fuzzing loop. Most fuzzers are developed based on AFL. AFLFast\cite{bohme2016coverage} models the process of generating new test cases as Markov chain and gravitates fuzzers towards low-frequency paths to increase coverage MOpt\cite{lyu2019mopt} uses particle swarm optimization to improve mutation strategies. AFL++\cite{fioraldi2020afl++} combines other open source fuzzers into a new fuzzer, which contains a variety of novel improvements. LibAFL\cite{fioraldi2022libafl} deconstructs AFL into modules to integrate orthogonal techniques.

\textbf{Directed White-box Fuzzing.} At first, directed fuzzing doesn't achieve by lightweight instrumentation. It is referred as directed white-box fuzzing, which mainly relies on symbolic execution and generates exploitable test cases\cite{ma2011directed}. However, the problem of path explosion and heavy constraint solving do harm to the scalability of directed symbolic execution\cite{stephens2016driller}. BugRedux\cite{jin2012bugredux} takes a sequence of statements as input and then generate a test case to trigger the crash of PUT.

\textbf{Directed Grey-box Fuzzing.} According to the paradigm of problem, directed symbolic execution casts directness as an \textit{iteration} problem while directed grey-box fuzzing define distance as \textit{loss function} and further treat the reachability of test cases as an optimization problem. AFLGo\cite{bohme2017directed} is the first directed grey-box fuzzer. It realizes directness through power schedule based on simulated annealing. Hawkeye\cite{chen2018hawkeye} makes a comprehensive improvement (e.g., mutation strategy) on AFLGo. WindRanger\cite{du2022windranger} defines the Deviation Basic Block (DBB), which distinguishes important BBs and unimportant BBs. Based DBB, it makes a comprehensive improvement too. FuzzGuard\cite{zong2020fuzzguard} is an extension based on AFLGo. It uses a neural network to predict the reachability of a test case before execution and the test case is skipped if it is not reachable. BEACON\cite{huang2022beacon} utilizes backward interval analysis to insert several assertions before the branches of the program, so that the test cases that cannot reach the target are terminated in advance. One of major differences between BEACON and FuzzGuard is that BEACON cuts-the-loss in a provable way.

\section{Conclusion}

We presents FGo, which terminates unreachable test cases early with exponentially increasing probability. FGo is faster than AFLGo in reproducing crashes. Compared to other directed grey-box fuzzers, FGo makes full use of the unreachable information contained in iCFG and doesn‘t generate any additional overhead caused by reachability analysis. Moreover, this idea can be integrated into other fuzzers easily.

\section*{Acknowledgement}

\bibliographystyle{acm}
\bibliography{{fgo}}

\end{document}